\documentclass[10pt,sigplan,screen,nonacm]{acmart}

\usepackage{graphicx}
\usepackage{colortbl}
\usepackage{microtype}
\usepackage{balance}
\usepackage{mathtools}

\begin{document}

\title{ChiBench: a Benchmark Suite for Testing Electronic Design Automation Tools}

\author{Rafael Sumitani}
\orcid{0009-0003-5226-1966}
\affiliation{%
  \institution{UFMG}
  \city{}
  \country{Brazil}
}
\email{rafaelsumitani@dcc.ufmg.br}

\author{Jo\~{a}o Victor Amorim}
\orcid{0009-0001-6775-2579}
\affiliation{%
  \institution{UFMG}
  \city{}
  \country{Brazil}
}
\email{joao.amorim@dcc.ufmg.br}

\author{Augusto Mafra}
\orcid{0000-0000-0000-0000}
\affiliation{%
  \institution{Cadence}
  \city{}
  \country{Brazil}
}
\email{augusto@cadence.com}

\author{Mirlaine Crepalde}
\orcid{0000-0000-0000-0000}
\affiliation{%
  \institution{Cadence}
  \city{}
  \country{Brazil}
}
\email{mirlaine@cadence.com}

\author{Fernando M Quint\~{a}o Pereira}
\orcid{0000-0002-0375-1657}
\affiliation{%
  \institution{UFMG}
  \city{}
  \country{Brazil}
}
\email{fernando@dcc.ufmg.br}

\begin{abstract}
Electronic Design Automation (EDA) tools are software applications used by engineers in the design, development, simulation, and verification of electronic systems and integrated circuits.
These tools typically process specifications written in a Hardware Description Language (HDL), such as Verilog, SystemVerilog or VHDL.
Thus, effective testing of these tools requires benchmark suites written in these languages.
However, while there exist some open benchmark suites for these languages, they tend to consist of only a handful of specifications.
This paper, in contrast, presents ChiBench, a comprehensive suite comprising 50 thousand Verilog programs.
These programs were sourced from GitHub repositories and curated using Verible's syntactic analyzer and Jasper\textsuperscript{TM}'s HDL semantic analyzer.
Since its inception, ChiBench has already revealed bugs in public tools like Verible's obfuscator and parser.
In addition to explaining some of these case studies, this paper demonstrates how ChiBench can be used to evaluate the asymptotic complexity and code coverage of typical electronic design automation tools.
\end{abstract}

\ccsdesc[500]{Software and its engineering~Compilers}

\keywords{Benchmark, Verilog, Testing}

\maketitle

\section{Introduction}
\label{sec:intro}

EDA (Electronic Design Automation) tools are software applications used by engineers in the design, development, simulation, and verification of electronic systems and integrated circuits (ICs).
These tools cover various stages of the electronic design process, from conceptualization and design entry to implementation, verification, and testing.
Examples of EDA tools include Cadence Jasper for formal verification\footnote{Available at \url{https://www.cadence.com/en_US/home/tools/system-design-and-verification/formal-and-static-verification.html}} and Verible's tool\footnote{Available at \url{https://github.com/chipsalliance/verible}}. 
These tools operate on similar types of input data: programs in some Hardware Description Language (HDL), such as Verilog, SystemVerilog, VHDL or SystemC.
Thus, the effective development and testing of such tools require benchmarks in these languages.

\paragraph{Verilog Benchmarks (or their lack thereof)}
A {\it Benchmark Suite} is a collection of programs used to test computing systems that process such programs.
There exist open source benchmark collections tailored for EDA tools, such as the ISCAS Benchmark Circuits~\cite{Brglez89} (31 circuits), the MCNC Benchmark Circuits~\cite{Kozminski91} (19 circuits in the YAL format), the EPFL Combinational Benchmark Suite~\cite{Amaru19} (23 circuits), the RAW Benchmark Suite~\cite{Babb97} (twelve programs), the KOIOS collection (19 circuits implementing different neural networks) and the Titan23 suite of 23 circuits~\cite{Murray15}.
These collections contain a small number of programs: typically less than 50.
This fact is unfortunate because, in the works of \citet{Wang18}: ``{\it Although there are numerous benchmark sites publicly available, the number of programs available is relatively sparse compared to the number that a typical compiler will encounter in its lifetime.}''
This paper mitigates this problem, releasing a much larger collection of Verilog circuits.

\paragraph{The Contribution of this Work}
This paper describes ChiBench, an open collection of 50K Verilog programs, mined from open-source GitHub repositories.
These programs were curated in a three-step process, which
Section~\ref{sec:sol} explains:
first, Verilog codes were automatically scraped from public code repositories.
In a second phase, each program was parsed using Verible's syntactic analyzer, to ensure compliance with the IEEE Standard 1364~\cite{Thomas96}.
Finally, these programs were sieved via Jasper's HDL semantic analyzer, to ensure that each circuit contains all the required dependencies.

This paper illustrates three different usage scenarios where ChiBench has been employed, using two tools from the Verible project as test subjects.
First, Section~\ref{sub:complexity} demonstrates how ChiBench programs can be used to investigate the asymptotic complexity of EDA tools.
Second, Section~\ref{sub:coverage} explains how these programs can test EDA tools by gauging the coverage of ChiBench as a test dataset.
Finally, Section~\ref{sub:bugs} briefly describes two bugs discovered in open-source tools through ChiBench-based tests.

\section{The Construction of a Benchmark Suite}
\label{sec:sol}

In order to build our Benchmark Suite, we have mined programs from open-source GitHub repositories, using
GitHub's REST API \footnote{Available at \url{https://docs.github.com/en/rest}}. We
use GitHub's API to build a list of candidate Verilog repositories.
Said list is sorted by popularity (measured as the number of stargazers).
We remove from the candidate list repositories that are not available for public usage, due to the lack of a license.
Thus, for each repository $R$ in the sorted list, we have implemented a Python script that proceeds as follows:

\begin{enumerate}
    \item Clone $R$ and locally copy all its \texttt{.v} files;
    \item Assigns a unique name to each \texttt{.v} file, based on its repository and its local path;
    \item Remove any special characters from the file's name to avoid encoding issues.
\end{enumerate}
We repeat the above sequence of steps for all the repositories in the base list,
until reaching a predefined number of files.
This threshold is set upon calling the mining script.
Currently, ChiBench provides only Verilog programs; however, the mining script can be easily adapted to fetch files in any other language that is tagged in GitHub, such as VHDL.

\subsection{Curating the Data}
\label{sub:curate}

After we have copied the necessary number of Verilog files from GitHub, we proceed to select valid programs.
To this effect, we only keep files that are syntactically and semantically valid.
Thus, this process involves passing the files through two sieves.
The first sieve, the syntax analysis, happens via the Verible syntactic analyzer.
At this stage, if Verible's parser cannot build an abstract syntax tree for a file, we discard it.
Example~\ref{ex:syntax} illustrates one such situation.

\begin{figure}[ht]
\centering
\includegraphics[width=\columnwidth]{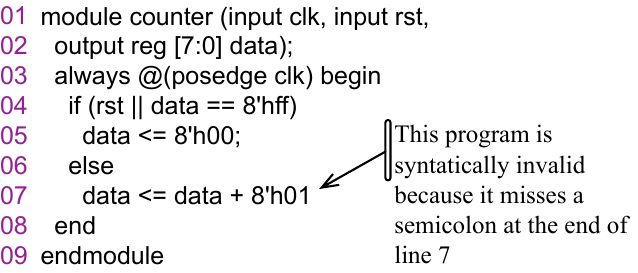}
\caption{Verilog specification filtered out by our syntactic verification.}
\Description{Verilog specification filtered out by our syntactic verification.}
\label{fig:syntactically_invalid_v}
\end{figure}

\begin{example}
\label{ex:syntax}
The program in Figure~\ref{fig:syntactically_invalid_v}, which specifies an 8-bit counter, will be filtered out by the syntactic filter.
It contains a missing semicolon at Line 7.
Such syntactically invalid files are uncommon in the mining process.
Nevertheless, they occur, as the repositories contain, for instance, files that are still under development.
\end{example}

Once we remove any syntactically invalid programs, we use Jasper's HDL semantic analyzer to filter out any semantically invalid programs\footnote{Jasper's HDL semantic analyzer is triggered with the \texttt{analyze} built-in
command.}.
Notice that Jasper's HDL analyzer also rejects invalid syntax.
However, Jasper's HDL analyzer is more computationally expensive than Verible's because it also considers semantic analysis, being more restricted to Verilog language standards. Consequently, in order to reduce the number of programs sent for semantic analysis, we chose to filter out syntactically invalid programs before using Jasper.
Example~\ref{ex:semantic} better explains what is the role of the semantic analysis.

\begin{figure}[ht]
\centering
\includegraphics[width=\columnwidth]{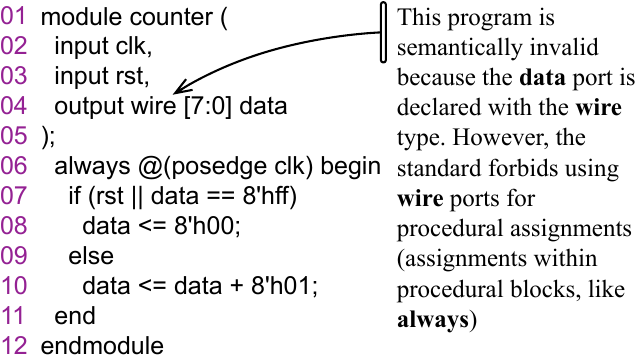}
\caption{Verilog specification that fails the semantic test.}
\Description{Verilog specification filtered out by our semantic verification.}
\label{fig:semantically_invalid_v}
\end{figure}

\begin{example}
\label{ex:semantic}
Figure~\ref{fig:semantically_invalid_v} shows an example of a program that fails the semantic sieve due to a type inconsistency.
In this case, the IEEE standard forbids the declaration of data ports with the wire type.
Our data generation process considers each file independently, thus, this error might occur.
Nevertheless, our experience is that most Verilog programs can be successfully validated as a single compilation unit.
\end{example}

\subsection{Licensing}
\label{sub:license}

\begin{figure}[ht]
\centering
\includegraphics[width=\columnwidth]{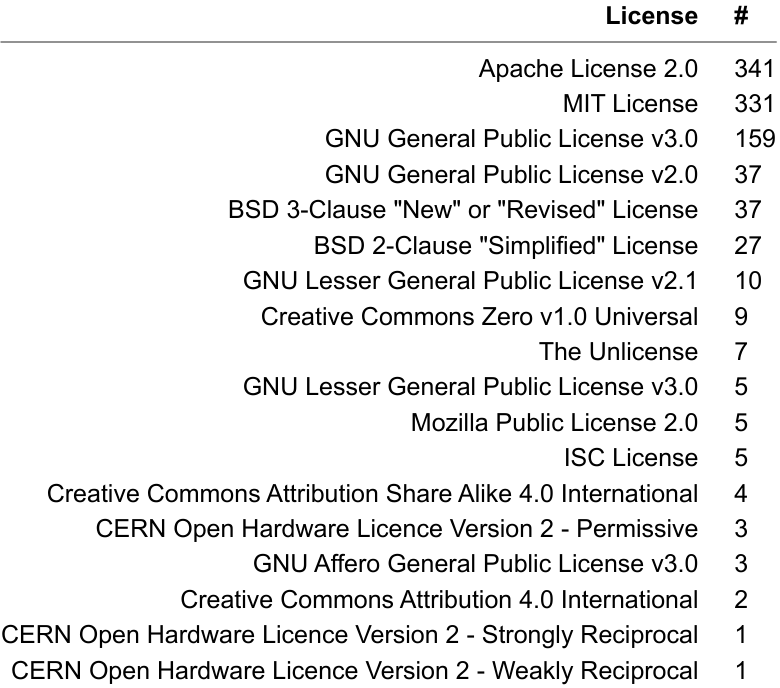}
\caption{Licenses of repositories used to build ChiBench.}
\Description{Licenses of repositories used to build ChiBench.}
\label{fig:licenses}
\end{figure}

ChiBench only contains files that provide permissive licenses.
In other words, we remove from the public distribution of this collection any Verilog specification that comes from either a repository without a license or that comes from a repository whose license prevents distribution.
Figure~\ref{fig:licenses} shows the number of licenses in each repository used to build ChiBench.
Each ChiBench file contains, as a comment, the URL of the repository from where it comes, plus its license.

\section{Evaluation}
\label{sec:eval}

The goal of this section is to demonstrate that ChiBench is a useful collection of benchmarks.
To this end, we shall investigate the following three research questions:
\begin{description}
\item[RQ1:] Can ChiBench Programs be used to infer, empirically, the asymptotic complexity of EDA tools?
\item[RQ2:] How much coverage should we expect to obtain by using ChiBench as a dataset to test EDA tools?
\item[RQ3:] Are ChiBench-based tests able to uncover zero-day bugs in well-known EDA tools?
\item[RQ4:] How are ChiBench's programs characterized in terms of size?
\end{description}

\paragraph{Experimental Setup}
The evaluation described in this Section uses two tools available in the Verible Project (Release v0.0-3622-g07b310a3, Mar 13th, 2024) as test subjects: the parser and the obfuscator.
Experiments run on an AMD Ryzen 9 5900X 12-Core 3.7GHz processor featuring Ubuntu Linux 22.04.4 LTS (kernel 6.5.0-27-generic).
Coverage is measured via Clang's source-based code coverage feature (available in Clang 14.0.0).

\subsection{RQ1 -- Asymptotic Analysis}
\label{sub:complexity}

The asymptotic complexity of a tool is an expression that relates the running time of that tool as a function of its input size.
Determining an analytical formula for the asymptotic complexity of a tool is a challenging problem, as this formula is influenced by many details of that tool's implementation.
Thus, as an alternative, to understand the asymptotic behavior of implementations of algorithms, researchers resort to the so-called {\it empirical complexity analysis}.
In the words of \citet{Sumitani23}:
``{\it Empirical Complexity Analysis is a branch of computer science that tries to infer the asymptotic complexity of algorithms through the observation of multiple executions of that algorithm with different inputs}''.
In this paper, we show how ChiBench can be effectively used as a means to carry out empirical complexity analysis.

\begin{figure}[ht]
\centering
\includegraphics[width=\columnwidth]{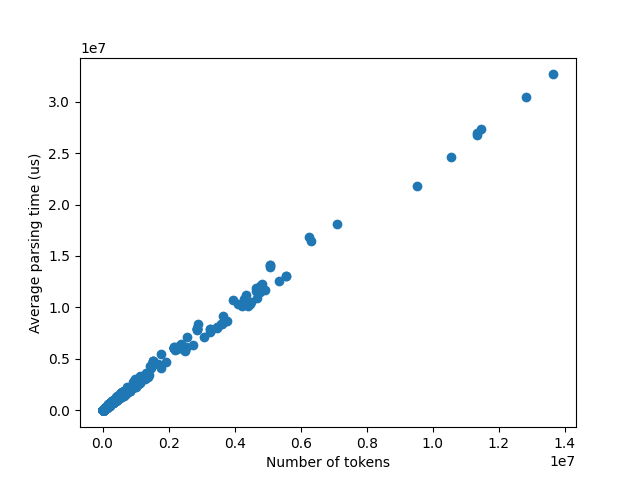}
\caption{The impact of the number of tokens on Verible's parsing time.
This figure contains about 50K points---one point per program in the ChiBench collection.}
\Description{The impact of the number of tokens on Verible's parsing time.}
\label{fig:complexity}
\end{figure}

\paragraph{Discussion}
Figure~\ref{fig:complexity} relates the number of tokens in ChiBench programs with the time that Verible's syntactic analyzer takes to parse these programs.
To perform this experiment, we used the programs in the ChiBench collection as input to Verible's parser.
As Figure~\ref{fig:complexity} shows, the running time behavior of Verible is linear.
In this regard, Pearson's coefficient relating running time and number of tokens is 0.99, with a p-value of less than $2^{-16}$. Therefore, Verible's parser is expected to run in linear time on the number of tokens that form the program with very strong probability.

\subsection{RQ2 -- Coverage Analysis}
\label{sub:coverage}

{\it Coverage} is a metric that quantifies the effectiveness of a test suite.
In this paper, we report coverage in two ways.
Either as the number of lines in the source code of a program that the test suite exercises.
Or else as the number of branches exercised by the test case in the binary representation of that same program.
Thus, we define the coverage ratio as either the number of lines covered divided by the total lines of code or as the number of branches covered divided by the total number of branches.
The higher the coverage ratio, the better the test suite.
This section analyzes the coverage ratio of ChiBench.

\begin{figure*}[ht]
\centering
\includegraphics[width=\textwidth]{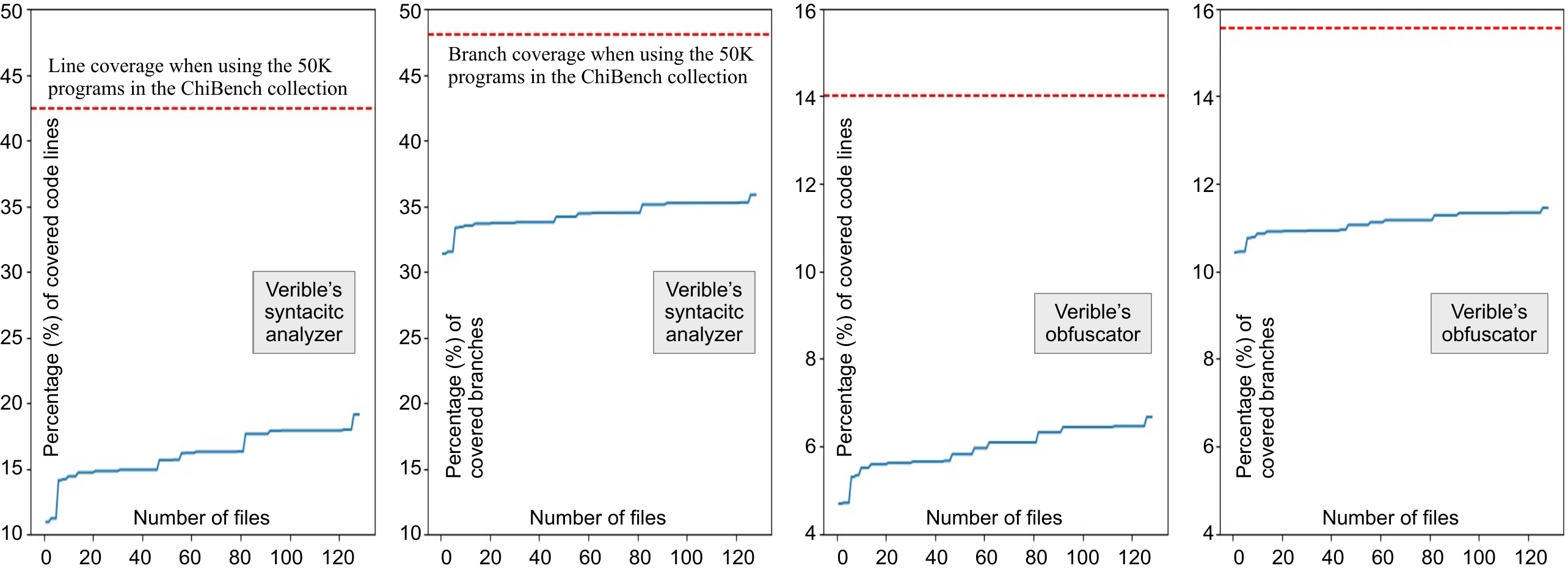}
\caption{Code coverage of different tools of the Verible Framework, using the 128 largest programs from ChiBench.}
\Description{Code coverage of different tools of the Verible Framework, using the 128 largest programs from ChiBench.}
\label{fig:codeCoverage}
\end{figure*}

\paragraph{Discussion}
Figure~\ref{fig:codeCoverage} presents the code coverage for Verible's obfuscator and parser assessed with the 128 largest programs from ChiBench.
The coverage pattern for both tools is similar; however, the parser demonstrates a notably higher coverage. For lines covered it ranges from 10\% to 19\% against 4.5\% to 7\% for the obfuscator. In terms of branches covered, the parser starts at 31\% and reaches 36\% whereas the obfuscator begins at 11\% and ends at nearly 12\%.
The red lines represent the code coverage when using all programs from ChiBench.
For the parser, we achieved 42\% coverage for lines and 48\% for branches.
In comparison, the obfuscator reached 14\% for lines and nearly 16\% for branches.

We hypothesize that this disparity arises because the parser is a larger feature and, consequently, calls a larger variety of functions from Verible's modules. In contrast, the obfuscator is a more self-contained tool.
Although these numbers seem low in principle, we remind the reader that Verible is a large framework, which includes several tools, such as a linter, a code formatter, and a language server.
The code that forms these parts, although part of the Verible binary library, remained largely unseen during this experiment.

\subsection{RQ3 -- Actual Bug Reports}
\label{sub:bugs}

One of the primary goals of a benchmark suite is to uncover bugs in the implementation of tools that read this suite.
ChiBench was initially conceived to stress-test any EDA tool and to demonstrate the effectiveness of this collection as a bug-finding mechanism, we apply it onto open-source tools available in the Verible Framework.

\paragraph{Discussion}
We have used ChiBench to evaluate the correctness of Verible's parser and code obfuscator.
In this case, we define as a {\it buggy evaluation} the analysis of a program from ChiBench that results in a ``crash''; that is, an execution of the subject tool that leads to either an assertion or to an operating system exception (such as a segmentation fault).
Under these definitions, we have observed one buggy evaluation on the parser and another on the obfuscator.
Both have been reported to the Verible's community:
\begin{itemize}
    \item \url{https://github.com/chipsalliance/verible/issues/2159}: \\
    Verible's obfuscator crashes when reading a program that only contains the pragma directive.
    \item \url{https://github.com/chipsalliance/verible/issues/2181}: \\
    Verible's parser crashes instead of reporting syntax errors related to instantiation type.
\end{itemize}
Interestingly, the second issue---which has been acknowledged as a true bug---was not activated by a program in ChiBench itself, but by a program derived from the suite: to maximize the diversity of ChiBench programs, we have used this collection to generate new programs.
Generation works as follows: we use the 50K programs in ChiBench to calculate the probability that each production rule in Verible's parser is exercised when parsing a program.
Then, we use this probabilistic grammar to generate new programs.
In this case, every type is set as \texttt{logic}.

\subsection{RQ4 -- Size Characterization}
\label{sub:size}

ChiBench programs are mined from open-source repositories; hence, we speculate that they approximate the average Verilog program that represents typical circuits.
This section provides some characterization of such programs.
To this end, we analyze their size distribution.

\begin{figure}[ht]
\centering
\includegraphics[width=\columnwidth]{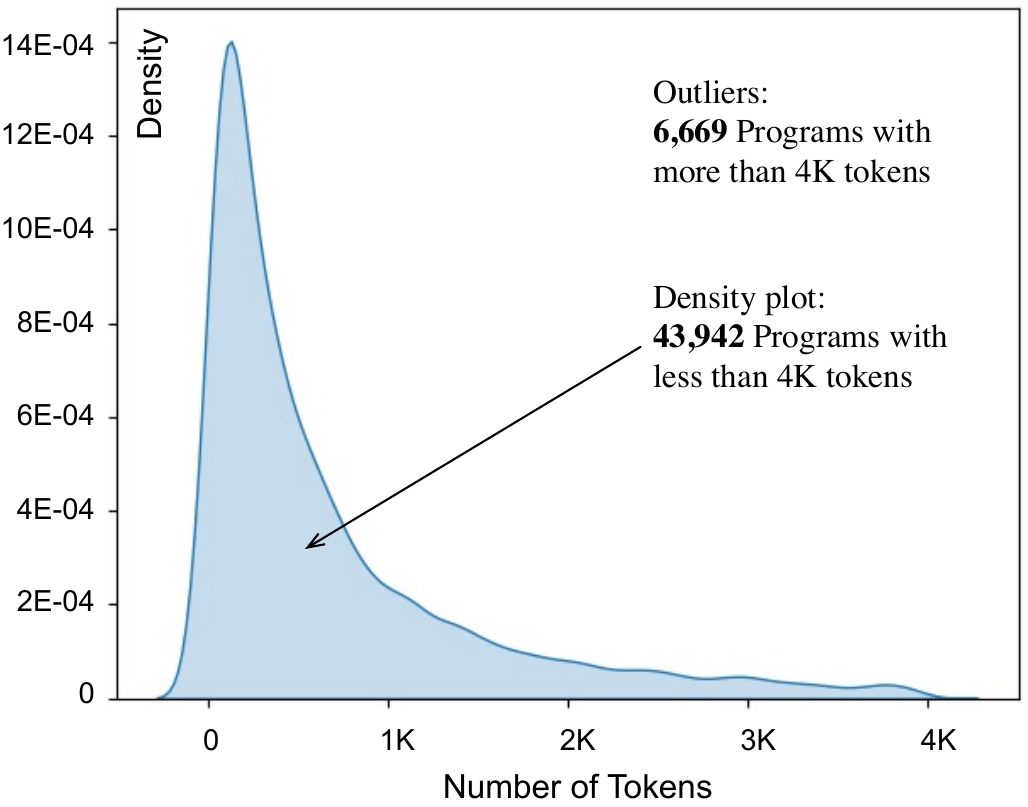}
\caption{Distribution of ChiBench programs per size, measured in number of tokens.}
\Description{Distribution of ChiBench programs per size, measured in number of tokens.}
\label{fig:densityCurve}
\end{figure}

\paragraph{Discussion}
Figure~\ref{fig:densityCurve} shows a density curve representing the size distribution of ChiBench programs.
We measure size as the number of tokens that the Verible lexer produces for each Verilog file.
Notice that this metric does not depend on user-defined names.
Figure~\ref{fig:densityCurve} makes it clear that most Verilog specifications are small.
In total, ChiBench contains 50,611 Verilog programs.
Out of this lot, 66.87\% contain less than 1,000 tokens, and 79.24\% contain less than 2,000.
The largest program in ChiBench contains 25,690,281 tokens, and the smallest contains only 4.
The average number of tokens is 33,350.9, and the median number is 489.

\section{Related Work}
\label{sec:rw}

As we have already mentioned in Section~\ref{sec:intro}, there exist already collections of benchmarks formed by hardware specification languages~\cite{Brglez89, Kozminski91, Amaru19, Babb97, Murray15}.
However, these collections are small: never containing more than 50 circuits.
Nevertheless, with the rising popularity of large language models, this scenario has seen changes in the last year.

As an example of this new trend, at the end of 2023, \citet{Thakur24} released VeriGen, a version of the CodeGen~\cite{Nijkamp23} fine-tuned for the synthesis of Verilog specifications.
In the process of tuning CodeGen, \citeauthor{Thakur24} have collected 50K Verilog circuits.
However, in contrast to ChiBench, the dataset used by \citeauthor{Thakur24} has not undergone any form of filtering; hence, we do not know if these programs are semantically valid.
This dataset is publicly available\footnote{At \url{https://huggingface.co/datasets/shailja/Verilog_GitHub}}; however, each program is a single line string.
Parsing these programs automatically to reconstruct the original Verilog specification is not possible, due to the presence of line comments in the codes.
This shortcoming is not a problem in the context of \citeauthor{Thakur24}'s work, given that they are interested in building a large language model that is based on k-grams of Verilog codes.
Yet, the dataset used to train the model could not be used, for instance, to stress-test EDA tools.
We believe that ChiBench might be useful to support the implementation of models like VeriGen.
To this effect, independent evaluations of VeriGen have found that ``{\it A primary
contributing factor to this shortfall [the inability to uncover bugs in EDA tool] is the insufficiency of HDL code resources for training}''~\cite{Nijkamp23}.
This perceived lack of benchmarks seems to be a common issue among researchers working on large language models for hardware specifications~\cite{Yao24, Tsai24}.

\section{Conclusion}
\label{sec:con}

This paper has described ChiBench, a collection of 50K Verilog programs mined from open-source GitHub repositories.
In addition to explaining the methodology to build this suite, this paper showed how ChiBench can be effectively used to test and analyze the behavior of electronic design automation tools.
The infrastructure used in the construction of ChiBench can be adjusted to other languages, such as VHDL.
Thus, future work might involve maximizing the diversity of ChiBench programs to broaden coverage of EDA tools.

\paragraph{Software}
ChiBench is available at \url{https://github.com/lac-dcc/chimera}.
Each benchmark available in ChiBench provides, as a header comment, its original license, in addition to the repository from where that specification was obtained.

\section*{Acknowledgement}
This project is sponsored by Cadence Design Systems.
Additionally, the authors acknowledge the support of CNPq (grant 406377/2018-9), FAPEMIG (grant PPM-00333-18), and CAPES (grant ``Edital CAPES PrInt'').

\bibliographystyle{ACM-Reference-Format}

\end{document}